\documentclass[10pt,groupedaddress,prb,twocolumn]{revtex4}

\usepackage{amssymb}
\usepackage{amsmath}
\usepackage{epsfig}

\begin{document}



\title{Nanowire metamaterials with extreme optical anisotropy}

\author{Justin Elser}
\email{elserj@physics.oregonstate.edu}

\author{Robyn Wangberg }

\author{
Viktor A. Podolskiy}
\email{viktor.podolskiy@physics.oregonstate.edu}
\homepage{http://www.physics.oregonstate.edu/~vpodolsk}

\affiliation{
Physics Department, 
301 Weniger Hall, Oregon State University, 
Corvallis OR 97331}

\author{Evgenii E. Narimanov}
\affiliation{
EE Department, Princeton University, 
Princeton NJ 08540}



\begin{abstract}
We study perspectives of nanowire metamaterials for negative-refraction waveguides, high-performance polarizers, and polarization-sensitive biosensors. We demonstrate that the behavior of these composites is strongly influenced by the concentration, distribution, and geometry of the nanowires, derive an analytical description of electromagnetism in anisotropic nanowire-based metamaterials, and explore the limitations of our approach via three-dimensional numerical simulations. Finally, we illustrate the developed approach on the examples of nanowire-based high energy-density waveguides and non-magnetic negative index imaging systems with far-field resolution of one-sixth of vacuum wavelength. 
\end{abstract}
\maketitle

The anisotropy of effective dielectric permittivity is widely used in optical, infrared (IR), THz and GHz sensing, spectroscopy, and microscopy\cite{anisot4,belov,shvetsPRL,shultzTHzpolarizer}. Strongly anisotropic optical materials can be utilized in non-magnetic, non-resonant optical media with negative index of refraction, and have the potential to perform subdiffraction imaging and to compress the radiation to subwavelength areas\cite{enghetaWG,belov,podolskiyPRB,govyadinovFunnels}. The performance of these polarization-sensitive applications can be related to the relative difference of the dielectric constant along the different directions. In the majority of natural anisotropic crystals this parameter is below $30\%$ \cite{palik}. While it may be sufficient for some applications, a number of exciting phenomena ranging from high-performance polarization control\cite{shultzTHzpolarizer} to subwavelength light guiding\cite{govyadinovFunnels,enghetaWG,belov} to planar imaging\cite{podolskiyPRB} require different components of a permittivity tensor to be of {\it different signs}. 

In this Letter we study the perspectives of using nanowire composites as {\it meta-materials} with extreme optical anisotropy. We demonstrate that even $10\%$ stretching/compression of the nanowire structures may dramatically affect the electromagnetic properties of these systems and {\it change the sign of components of the permittivity tensor}. We present an analytical description of wave propagation in anisotropic nanowire composites -- Generalized Maxwell-Garnett approach (GMG), and verify our technique via three-dimensional (3D) numerical simulations. Finally, we illustrate our approach on the examples of several nanowire-based systems for light compression below the diffraction limit and negative refraction-imaging with far-field resolution of $\lambda_0/6$ (with $\lambda_0$ being free-space wavelength).

The use of metallic wire mesh as anisotropic low-frequency plasma has been proposed in\cite{pendryWires} and experimentally realized for normal light incidence in\cite{shultzTHzpolarizer,shultzNIM}. However, the applicability of these nanowire-based materials for any non-trivial geometry involving oblique light incidence or wave-guiding is still considered to be questionable due to strong nonlocal interactions\cite{shvetsWires}, that may potentially result in {\it positive} components of the permittivity tensor. Furthermore, the majority of existing effective-medium theories (EMTs)\cite{shvetsWires,MG,stroud,miltonBook} are limited to the optical response of nanowires that are isotropically distributed in the host material. The predicted response of these systems is almost independent of nanowire distribution and is described by a single parameter -- nanowire concentration. These existing techniques are therefore not applicable for practical composites where the geometry is anisotropic due to fabrication process or as a result of a controlled mechanical deformation\cite{park}. Understanding the optical behavior of nanowire structures beyond one-parameter EMT is the main purpose of this Letter. 

\begin{figure}[th] 
\centerline{\epsfig{file=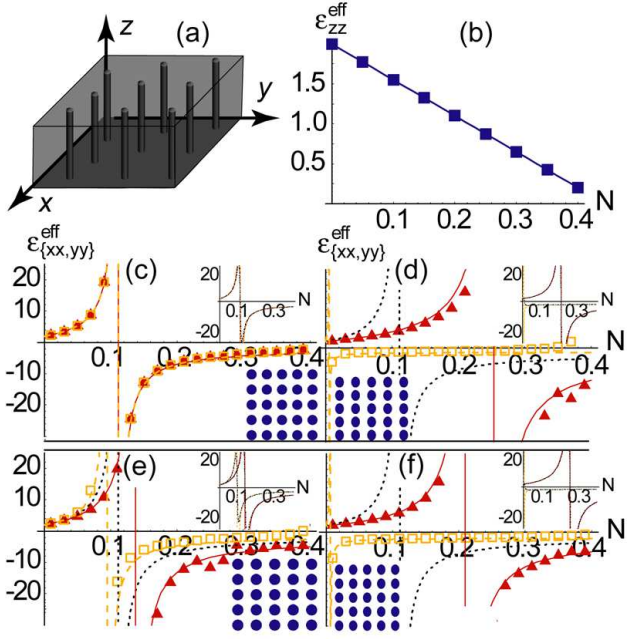,width=3.25in}}
\vspace{10pt}
\caption{
(color online) (a) Schematic geometry of a nanowire composite. (b) $\epsilon^{\rm eff}_{zz}$ for $\epsilon^{\rm in}=-2.5$, $\epsilon^{\rm out}=2$ (Ag nanowires in a polymer for $\lambda_0\simeq 360nm$). (c-e) $\epsilon^{\rm eff}_{xx}$ (red triangles, solid lines) and $\epsilon^{\rm eff}_{yy}$ (orange rectangles, dashed lines) for the composite in (a), with $\Lambda_{x}=\Omega_{x}=0$ (c), $\Lambda_{x}=0.2; \Omega_{x}=-0.2$ (d), $\Lambda_{x}=0.2; \Omega_{x}=0$ (e), $\Lambda_{x}=0; \Omega_{x}=-0.2$ (f); quasistatic numerical calculations (symbols); and GMG (lines); dotted lines in (d)$\cdots$(f) are identical to lines in (c). Bottom insets show crossections of composites for $N=0.35$; Black dashed lines in top insets illustrate the effect of small losses in nanowires ($\epsilon^{\rm in}=-2.5+0.04i$); note that this effect is negligible for $|\epsilon^{\rm eff}|\lesssim 20$. The breakdown of GMG occurs when the local field becomes inhomogeneous on the scale of $r_\alpha$}
\label{figEMT}
\end{figure}

The geometry of the nanowire composites considered in this work is shown in Fig.\ref{figEMT}. The nanowires with permittivity $\epsilon^{\rm in}$ are embedded into a host material with permittivity $\epsilon^{\rm out}$. The nanowires are aligned along the $z$ direction of Cartesian coordinate system. We assume that the nanowires have elliptic crossections with the semi-axes $r_x$ and $r_y$ directed along $x$ and $y$ coordinate axes respectively. We further assume that the homogeneous nanowire composite may be compressed or stretched, leading to the anisotropic distribution of individual nanowires. The typical separations between the nanowires $x$ and $y$ directions are denoted by $l_x$ and $l_y$\cite{footnoteLXLY}. In this work we focus on the case of ``homogeneous metamaterial'', when the inhomogeneity scale is smaller than the wavelength ($r_\alpha,l_\alpha \ll \lambda_0$) and nanowire concentration $N$ is small so that the lattice feedback effects can be treated using perturbative techniques as described below [for the case of $\epsilon^{\rm in}<0$ additional requirement $r_\alpha< \sigma$, with $\sigma$ being skin depth in wires, must be fulfilled]. Under these conditions, Maxwell equations have free-space-like solutions that can be represented as a series of plane electromagnetic waves propagating inside a material with some effective dielectric permittivity\cite{landauECM} $\epsilon^{\rm eff}$: 
\begin{equation}
\label{eqEpsEff}
 <D_\alpha>=\epsilon^{\rm eff}_{\alpha\beta}{<E_\beta>}.
\end{equation}
The angular brackets in Eq.(\ref{eqEpsEff})  denote the average over microscopically large (multi-wire), macroscopically small (subwavelength) region of the space, with Greek indices corresponding to Cartesian components, and assumed summation over the repeated indices. If both $\epsilon^{\rm in}$ and $\epsilon^{\rm out}$ are isotropic, the normal axes of the tensor of effective dielectric permittivity will coincide with the coordinate axes. Thus, in the selected geometry the permittivity tensor becomes diagonal: $\epsilon^{\rm eff}_{\alpha\beta}=\delta_{\alpha\beta}\epsilon^{\rm eff}_{\beta\beta}$, with $\delta_{\alpha\beta}$ being the Kronecker delta function. 

We now derive the expressions for the components of the effective permittivity $\epsilon^{\rm eff}_{xx},\epsilon^{\rm eff}_{yy}$ and $\epsilon^{\rm eff}_{zz}$. Using the continuity of the $E_z$ component, Eq.(\ref{eqEpsEff}) yields:
\begin{equation}
\label{eqEpsZ}
\epsilon^{\rm eff}_{zz}=N \epsilon^{\rm in}+(1-N)\epsilon^{\rm out}.
\end{equation} 
Note that similar to what was found in Refs.\cite{shvetsWires,MG,stroud,miltonBook}, the single parameter that determines the $z$ component of the permittivity in the effective medium regime is nanowire concentration $N$. 

To find ($x,y$) components of the $\epsilon^{\rm eff}$ we use the Maxwell-Garnett (MG) technique\cite{MG,stroud,miltonBook}. This approach assumes $N \ll 1$ so that the local field in the composite is homogeneous across a nanowire. The fields $D$ and $E$ are then averaged over a typical nanowire cell, and Eq.(\ref{eqEpsEff}) is used to extract the effective permittivity of a material. Naturally, average fields will have two contributions: one coming from the fields {\it inside} nanowires $E^{\rm in}$, and the second one coming from the fields {\it between} nanowires $E^{\rm out}$. The derivation of an EMT is therefore equivalent to understanding the relationship between $E^{\rm in}, E^{\rm out}$ and the {\it external} field acting on the system $E^0$. Conventional MG approach assumes that $E^{\rm out}=E^0$\cite{stroud,MG}, which is true only for the case of isotropic nanowire distributions. The crucial point of this work is that both $ E^{\rm in}$ and $ E^{\rm out}$ are strongly influenced by the {\it nanowire distribution} (given by the parameters $l_x,l_y$), and nanowire shape (described by $r_x,r_y$) along with nanowire concentration $N$.

We now derive the analytical expressions for $E^{\rm in}$ and $E^{\rm out}$. The typical excitation field acting on a nanowire in the sample will contain the major contribution from external field $E^0$ and the feedback field scattered by all other nanowires in the structure $\hat{\chi} E^0$, resulting in the effective excitation $E^0+\sum_j \hat{\chi}^j E^0=[\delta_{\alpha\beta}-\chi_{\alpha\beta}]^{-1} E^0_{\beta}$. For rectangular, triangular, and other highly-symmetrical lattices, as well as for a wide-class of random nanowire micro-arrangements, the feedback tensor becomes diagonal\cite{jacksonBook}, with the effective field acting on a nanowire being $[1-\chi_{\alpha\alpha}]^{-1}E^0_{\alpha}$\cite{footnoteIntCoef}.

Using the dimensionless function $S(\xi)= \sum_{ij}^\prime \frac{i^2}{( i^2+\xi^2 j^2)^2}$ with summation going over all pairs of $i$, $j$ except coordinate origin, the summation of $2D$ dipole fields over rectangular lattice shown in Fig.\ref{figEMT}(a) yields \cite{footnoteInfSum}
\begin{widetext}
\begin{eqnarray}
\chi_{\alpha\alpha}&=&
\frac{ ( {\epsilon_{\rm in}-\epsilon_{\rm out}})
r_x r_y P_\alpha }{4 l_x l_y}
\left[(\Lambda_\alpha+1)S(\Lambda_{\alpha}+1)-\frac{1}{\Lambda_\alpha+1}S\left(\frac{1}{\Lambda_{\alpha}+1}\right)\right]\nonumber
\\
& \simeq & 
- {0.16\; N\;\Lambda_{\alpha} P_\alpha (\epsilon^{\rm in}-\epsilon^{\rm out})}
\label{eqChi}
\end{eqnarray}
\end{widetext}
where we introduced the lattice distortion vector $\{\Lambda_{x},\Lambda_{y}\}=\{l_x/l_y-1,l_y/l_x-1\}$, and polarization term $P_\alpha=1/[{\epsilon^{\rm out}+ n_\alpha(\epsilon^{\rm in}-\epsilon^{\rm out})}]$, with $\{n_x,n_y\}=\{r_y/(r_x+r_y), r_x/(r_x+r_y)\}$ being the depolarization factors\cite{landauECM,miltonBook}. Note that the feedback parameter vanishes only for isotropic nanowire distribution $l_x=l_y$, corresponding to the well-known MG result\cite{MG,stroud,miltonBook,pendryWires,shvetsWires}.

This inter-wire interaction changes the ``microscopic'' field acting on the individual nanowires, and thus it directly affects both (homogeneous) field inside the nanowire $E^{\rm in}$\cite{landauECM}, 
\begin{eqnarray}
\label{eqEin}
E^{\rm in}_{ \alpha}&=&\frac{\epsilon^{\rm out}P_\alpha}
{1-\chi_{\alpha\alpha}}E^0_{\alpha}, 
\end{eqnarray} 
and the average field in-between the nanowires $E^{\rm out}$. Direct calculation of the average dipole\cite{landauECM} $E^{\rm out}$ of a given inclusion over the typical meta-material cell yields: 
\begin{widetext}
\begin{eqnarray}
\label{eqEout}
E^{\rm out}_{\alpha}&\simeq& \left[1+
\frac{N\;P_\alpha(\epsilon^{\rm in}-\epsilon^{\rm out})\left(Q(N) \cdot(
\Omega_{\alpha}+\Lambda_{\alpha})
-\pi\;\Omega_{\alpha}
\right)
}{
2\pi(1-N)(1-\chi_{\alpha\alpha})} 
\right] E^0_{\alpha},
\end{eqnarray} 
\end{widetext}
with $Q(N)= \pi-1-N(\pi-2)$ and shape vector $\{\Omega_{x},\Omega_{y}\}=\{r_x/r_y-1,r_y/r_x-1\}$. 

Combining Eqs.(\ref{eqEpsEff},\ref{eqEin},\ref{eqEout}) we arrive to the following expression for the in-plane components of permittivity in GMG approach:
\begin{equation}
\label{eqEpsXY}
\epsilon^{\rm eff}_{\alpha\alpha}=\frac{N\epsilon^{\rm in}E^{\rm in}_{\alpha}+
(1-N)\epsilon^{\rm out} E^{\rm out}_{\alpha}}
{N E^{\rm in}_{\alpha }+(1-N) E^{\rm out}_{\alpha }}.
\end{equation}

\begin{figure}[bth] 
\centerline{\epsfig{file=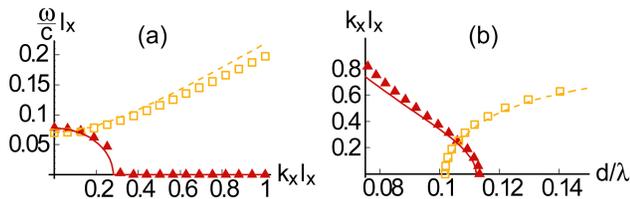,width=3.25in}}
\vspace{10pt}
\caption{
(color online) (a) dispersion of the fundamental TM (red triangles, solid lines) and TE (orange squares, dashed lines) modes in a waveguide with $d=400 nm$ with nanowire composite core; $\epsilon^{\rm in}=13; \epsilon^{\rm out}=-120; \Lambda_{x}=0.2; \Omega_{x}=-0.2; \l_x=40 nm; r_x=10 nm$; Note the negative refraction mode, predicted in \cite{podolskiyPRB}; (b) modal propagation constant for $\lambda_0=1.5 \mu m$ as a function of $d$; numerical solutions of 3D Maxwell equations (symbols); Eqs.(\ref{eqEpsZ},\ref{eqEpsXY},\ref{eqWave}) (lines). The breakdown of GMG correspond to $|k_\alpha l_\alpha|\sim 1$}
\label{figWave}
\end{figure}

To verify the accuracy of the developed GMG technique, we generate a set of nanowire composites with given values of $N$, $\epsilon^{\rm in}$, $\epsilon^{\rm out}$, $\Lambda$, and $\Omega$, excite each composite with a homogeneous field, and use the commercial finite-element partial differential equations solver, COMSOL Multiphysics 3.2\cite{COMSOL} to solve Maxwell equations, find the microscopic filed distribution, and calculate the average values of $E$, and $D$ over the volume of a composite, yielding $\epsilon^{\rm eff}$. In the simulations we used both random and periodic nanowire composites; the number of nanowires was sufficient ($>10^2$) to eliminate the dependence of $\epsilon^{\rm eff}$ on the sample size [finite-sample-size artifacts]. Fig.\ref{figEMT} shows the excellent agreement between GMG approach presented in this work and numerical solution of Maxwell equations in quasistatic limit for concentrations $N\lesssim 0.3$ and deformations $|\Omega_{\alpha}|, |\Lambda_{\alpha}|\lesssim 0.3$. Our simulations indicate that the quasi-static material properties are fully described by {\it average} parameters $(N, l_\alpha, r_\alpha)$. This particular property of the effective-medium composites indicates high tolerance of anisotropic metamaterials to possible fabrication defects. 

As expected, the field distribution across the nanowire structure and $\epsilon^{\rm eff}$ are strongly affected by $N$, as well as $\Lambda$ and $\Omega$. Thus, even $10\%$ anisotropy in inclusion shape or distribution may result {\it in change of sign of dielectric permittivity}. Such an effect opens the possibility to create optical materials with widely controlled opto-mechanical properties, potentially leading to new classes of polarizers, beam shapers, polarization-sensitive sensing and fluorescence studies, as well as for a wide class of ultra-compact waveguides\cite{enghetaWG,govyadinovFunnels} since the material properties may be tuned between $\epsilon\approx 0$ and $|\epsilon|\gg 1$. Some of these applications are described below. 

As it has been noted for GHz systems in\cite{shvetsWires}, the components of $\epsilon^{\rm eff}$ may be strongly affected by the spatial dispersion. To clarify these effects we used COMSOL package to identify the eigen waves propagating in $x$-direction through a planar waveguide with a composite core consisting of a rectangular array of $10\%$ Ag nanowires in Si host, extending from $z=0$ to $z=d$ (see Fig.1a), bounded by perfectly conducting walls (see \cite{podolskiyPRB} for the detailed explanation of the effects of waveguide walls and material absorption on the mode propagation). In Fig.\ref{figWave}(a) we show the agreement of the results of numerical solutions of 3D wave equations with the EMT dynamics of TE and TM modes propagating in a waveguide with homogeneous anisotropic core, described by
\begin{eqnarray}
\frac{\pi^2}{ \epsilon^{\rm eff}_{yy}\; d^2}+
\frac{k_x^{(TE)^2}}{\epsilon^{\rm eff}_{yy}}
=\frac{\omega^2}{c^2}; \;
\frac{\pi^2}{ \epsilon^{\rm eff}_{xx}\; d^2}+
\frac{k_x^{(TM)^2}}{\epsilon^{\rm eff}_{zz}}
=\frac{\omega^2}{c^2},
\label{eqWave}
\end{eqnarray}
with $\omega=2\pi/\lambda_0$ and $k_x, c, d$ being the modal wavevector, speed of light in the vacuum, and waveguide thickness, respectively. Note that this system does not support TEM modes\cite{podolskiyPRB}. 

\begin{figure}[tbh] 
\centerline{\epsfig{file=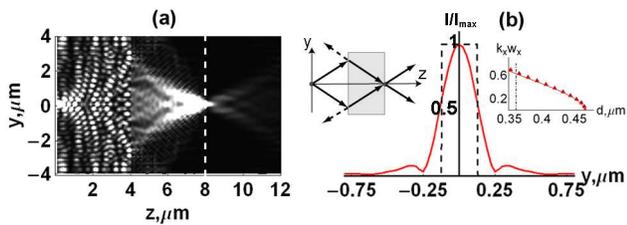,width=3.25in}}
\caption{
(color online) Planar waveguide imaging via nanowire materials $\lambda_0=1.5 \mu m; \; d=360 nm$; $n>0$ material: $\epsilon=13$; nanowire composite ($4<z\leq 8$): $r_x=r_y=10nm; \;l_x=l_y=50nm; \;\epsilon^{\rm in}=-120; \; \epsilon^{\rm out}=2$. (a) intensity across the system; (b) intensity distribution across the focal plane; insets show planar lens geometry and dispersion of a negative-index mode; dash-dotted line corresponds to $d=360 nm$ (a).} 
\label{figImag}
\end{figure}

Fig.\ref{figWave}(b) illustrates one of the applications of nanowire-based optical composites, {\it high-energy-density waveguide} -- a subwavelength structure supporting propagating {\it volume} modes. It is important to point out that in contrast to uniaxial media, anisotropic ($\epsilon^{\rm eff}_{xx}\neq\epsilon^{\rm eff}_{yy}$) nanowire composites can simultaneously support both $n>0$-TE and $n<0$-TM waves ($n=k_x c/\omega$). Moreover, the in-plane anisotropy (induced, for example, by deformation) can be used as a controlling mechanism in nanoscale nanowire-based pulse-management devices. 

It is clearly seen that the propagation of these modes is adequately described by GMG technique when $|k_\alpha l_\alpha|\ll 1$. As expected, the material properties in EMT regime are independent of nanowire micro-arrangements (type of crystalline lattice), while the exact point of EMT breakup ($|k_{\alpha}l_\alpha|_{\rm max} $) depends on local geometry and is maximized for almost-rectangular lattices [assumed in derivation of Eq.(\ref{eqChi})]. Note that the real requirement for EMT applicability, $|k_{\alpha}l_\alpha|\ll 1$, is different from the commonly used criterion $l_\alpha\ll\lambda_0$. Indeed, our simulations show that the spatial dispersion leads to cut-off of the modes with $|k_{\alpha}l_\alpha|\gg 1$, similar to what has been predicted for GHz wire systems \cite{shvetsWires} and nanolayer-based photonic funnels\cite{govyadinovFunnels}. 

Another application of nanowire structures, non-magnetic negative-index materials \cite{podolskiyPRB} is illustrated in Fig.\ref{figImag}. It is seen that the nanowire materials may be used to achieve sub-diffraction ($\lambda_0/6$) far-field resolution in the planar-lens geometry. 

In conclusion, we have developed the effective-medium theory (GMG) that adequately describes the optical properties of nanowire composites with anisotropic crossections and arrangements. Limitations of the proposed approach have been studied via numerical modeling. We demonstrated that the nanowire composites can be used to achieve extreme anisotropy at optical and IR frequencies, with controlled effective permittivity ranging from $\epsilon \ll-1 $ to $\epsilon\approx0$ to $\epsilon\gg 1$ -- thus leading to practical implementations of high-energy-density waveguides\cite{govyadinovFunnels,enghetaWG}, novel polarization-sensitive detectors, and recently proposed non-magnetic negative index systems\cite{podolskiyPRB}. Finally, we note that the technique presented here can be readily applied to dielectric, plasmonic, and polar-wire composites at optical, IR, and THz frequencies, and can be further extended to the cases of non-aligned inclusions, anisotropic $\epsilon^{\rm in}$ and $\epsilon^{\rm out}$, and 3D composites similar to what have been done for isotropic-arrangement cases in\cite{miltonBook,stroud}. 

This research is partially supported by GRF (OSU), Petroleum Research Fund (ACS), and PRISM (Princeton).





\begin{thebibliography}{}
\bibitem{footnoteLXLY}Mathematically, the average area of the unit cell $A$ in random anisotropic composite is related to the average separation between neighboring elements along two orthogonal directions $l_{\tilde{x}}$ and $l_{\tilde{y}}$ through the metric tensor $\hat{g}_{\alpha \beta}$: $A=\sum_{\{\alpha,\beta\}=\{\tilde{x},\tilde{y}\}}\hat{g}_{\alpha\beta}l_\alpha l_\beta$. The tensor becomes diagonal in ``eigen'' axes $\{\tilde{x},\tilde{y}\}=\{x,y\}$, used as primary directions throughout the manuscript.
\bibitem{footnoteIntCoef}
The role of parameters $\hat\chi$ used in this work is similar to that of interaction coefficients introduced for Lorentz model in\cite{collinBook}. These two techniques provide identical results for isotropic (in $x,y$ plane) composites, while the Eq.(\ref{eqEpsXY}) tends to be more robust than Lorentz model when $|\Lambda|, |\Omega|\sim 1$
\bibitem{footnoteInfSum}
Note that while $S(\xi) \rightarrow\infty$ in the limit $\xi \to 1$, the expression $\xi S(\xi)-S(1/\xi)/\xi$ remains finite for $\xi=1+\Delta$ with $|\Delta|\ll 1$.
\end{thebibliography}
\end{document}